\documentclass[amsmath,amssymb,prb]{revtex4-2}
\usepackage{graphicx}
\usepackage{dcolumn}
\usepackage{bm}
\usepackage{xcolor}

\begin{document}

\preprint{APS/123-QED}

\title{Inverse design and optical vortex manipulation for thin-film absorption enhancement}

\author{Munseong Bae}
\affiliation{Department of Mechanical Engineering, Massachusetts Institute of Technology (MIT), Cambridge, MA 02139 USA}%
 \affiliation{Now, Department of Electronic Engineering, Hanyang University, Seoul, 04763, South Korea}

\author{Jaegang Jo}
 \affiliation{Department of Electronic Engineering, Hanyang University, Seoul, 04763, South Korea}

\author{Myunghoo Lee}
 \affiliation{Department of Physics, Hanyang University, Seoul, 04763, South Korea}
 \affiliation{Department of Electronic Engineering, Hanyang University, Seoul, 04763, South Korea}

\author{Joonho Kang}
 \affiliation{Department of Electronic Engineering, Hanyang University, Seoul, 04763, South Korea.}

\author{Svetlana V Boriskina$^*$}%
 \affiliation{
 Department of Mechanical Engineering, Massachusetts Institute of Technology (MIT), Cambridge, MA 02139 USA
}%
 \author{Haejun Chung$^*$}%
\affiliation{%
 Department of Electronic Engineering, Hanyang University, Seoul, 04763, South Korea}
 \affiliation{%
 Department of Artificial Intelligence, Hanyang University, Seoul, 04763, South Korea
}%



\begin{abstract}
\begin{description}
\item[Abstract]
\end{description}
Optical vortices (OVs) have rapidly varying spatial phase and optical energy that circulates around points or lines of zero optical intensity. Manipulation of OV offers innovative approaches for various fields, such as optical sensing, communication, and imaging. In this work, we demonstrate the correlation between OVs and absorption enhancement in two types of structures. First, we introduce a simple planar one-dimensional (1D) structure that manipulates OVs using two coherent light sources. The structure shows a maximum of 6.05-fold absorption gap depending on the presence of OVs. Even a slight difference in the incidence angle can influence the generation/annihilation of OVs, which implies the high sensitivity of angular light detection. Second, we apply inverse design to optimize two-dimensional (2D) perfect ultrathin absorbers. The optimized free-form structure achieves 99.90\% absorptance, and the fabricable grating structure achieves 97.85\% at 775 nm wavelength. To evaluate OV fields and their contribution to achieving absorption enhancement, we introduce a new parameter, OV circularity. The optimized structures generate numerous OVs with a maximum circularity of 95.37\% (free-form) and 96.14\% (grating), superior to our 1D structure. Our study reveals the role of high-circularity localized OVs in optimizing nano-structured absorbers and devices for optical sensing, optical communication, and many other applications.
\end{abstract}

\maketitle
\section{\label{sec:level1}Introduction}

The optical vortex (OV) field engineering is of high interest for the control of optical information transport and light-matter interactions~\cite{dennis2009singular,soskin2016singular}, with applications ranging from engineering OV beams carrying the orbital angular momentum (OAM) of light~\cite{berkhout2008method,bekshaev2011internal, yue2016vector,huang2020ultrafast} to tunable high-quality-factor OV-rich modes in plasmonic and optoplasmonic nanostructures~\cite{boriskina2012molding,ahn2012electromagnetic}. OV is a region of circulating optical power flow, which forms around a point or line in space with zero field and an undefined phase, called a phase singularity. The total variation of the phase of the light field around a singularity is 2$\pi i$, where $i$ is the OV topological charge~\cite{berry2004optical,kotlyar2020evolution}, an integer describing the number and direction of the local optical phase rotation. Propagating free-space optical beams featuring a phase singularity at the beam center exhibit helical wavefront rotation and carry the optical angular momentum, which can be used for optical communication channel multiplexing~\cite{dashti2006observation,ramachandran2013optical,lei2015massive}, transferred to materials to encode information or spin nanoparticles~\cite{sreekanth2018biosensing,prajapati2019observation}, and used to improve the resolution of optical imaging techniques~\cite{weber2021minsted,rubinsztein2016roadmap,yao2011orbital}.

On the other hand, harnessing singular phase points in the electromagnetic fields reflected from optical material interfaces has been shown to significantly increase optical sensor sensitivity ~\cite{kravets2013singular,sreekanth2018biosensing,tsurimaki2018topological}. The phase of the reflected light exhibits a singularity at the frequency where perfect absorption is observed, and these two physical quantities are connected by the Kramers-Kronig relations~\cite{lucarini2005kramers}. Singular-phase reflectance has been successfully utilized to engineer high-performance photonic sensors over the past two decades. However, only recently it has been demonstrated that perfect absorption is topological in nature~\cite{tsurimaki2018topological} and is accompanied by the formation/annihilation of phase vortices in either the reciprocal energy-momentum space~\cite{liu2023spectral,sakotic2021topological,guo2017topologically} or in a generalized problem parameter space~\cite{berkhout2019perfect}.

Finally, the generation and annihilation of real-space OVs in the near-fields of 2D photonic nanostructures offer superior control over the local nanoscale optical energy flow~\cite{boriskina2014singular} and enable engineering sensitive optical switches and sensors~\cite{chung2021inverse,boriskina2011adaptive}. In this context, control of the OV generation in the optical near-fields of flat material interfaces either by multiple-wave interference~\cite{wolter2009concerning,lembessis2009surface,dennis2012topological} or by an external magnetic field~\cite{kim2022spontaneous} may enable new applications of the singular optics. Properly designed two-dimensional (2D) optical metasurfaces offer an even higher level of control and modification of spatial phases with high transmission efficiency~\cite{wang2011high, xie2018ultra,liu2022visible,doi:10.1021/acsphotonics.2c01007}. The generation of propagating vortex beams by optical metasurfaces has been studied theoretically and experimentally~\cite{chong2015polarization,genevet2015holographic,sroor2020high}. However, a comprehensive theory and design strategies to generate OV-rich optical interference fields inside thin films and metasurfaces are missing, as well as the understanding of the role OVs can play in the engineering of light absorption in optically-thin absorbers.

In this work, we systematically investigate the effect of OV formation in the thin-film photonic absorbers and demonstrate that OV generation and annihilation can be instrumental in either achieving absorption enhancement or in providing a mechanism for detecting the incidence angle of propagating waves. First, we analyze a simple planar one-dimensional (1D) structure consisting of a silicon thin film (750 nm thick) on a glass substrate illuminated by a light interference field with a wavelength of 775 nm. Our calculations predict the illumination conditions for the generation and annihilation of OVs, which cause dramatic absorption changes within three degrees of the incidence angle perturbation. Then, we use inverse design, a large-area computational optimization method, to generate an optical metasurface that allows achieving perfect absorption in a thin Si film and demonstrate that high absorptance is driven by the generation of high density of OVs pinned to the near-field of the metasurface. 

Inverse design, also known as adjoint optimization, is an optimization algorithm that allows the designing of large-area photonic devices with high-quality Figures of Merit (FoM), which can be uniquely defined for each application case~\cite{miller2012photonic,molesky2018inverse}. The inverse design has been successfully used to engineer and optimize optical metasurfaces~\cite{molesky2018inverse}, solar cells~\cite{miller2012photonic, ganapati2013light,molesky2018inverse}, optical resonators~\cite{lu2011inverse, molesky2018inverse, hoj2021ultra}, photonic bio-sensors~\cite{chung2021inverse,chung2022inverse}, and quantum photonic devices~\cite{dory2019inverse, yang2023inverse, otte2023tunable}. In this work, for the first time to the best of our knowledge, we apply the inverse design to model and explain the mechanism of the thin film perfect absorption with real-space near-field OV generation. The optimal inverse-designed devices exhibit perfect absorption with densely packed OVs in an ultrathin silicon film (of 100 nm thickness) at 775 nm wavelength. Specifically, a free-form OV-pinning metasurface without any design constraints except the minimum spatial feature size enables 99.90\% absorptance in the underlying ultrathin planar Si absorber, and the optimized lithography-compatible metasurface grating structure enables 97.85\% absorption in the same thin film. Both designs offer a dramatic performance improvement compared to a single-pass thin-film absorptance (5.14\%), a metric denoting the fraction of light absorbed during a single traversal through the material. 

\section{Angle-selective optical vortex manipulation in thin film}
\begin{figure*}[!bp]
\centering
\includegraphics[width=0.95\linewidth]{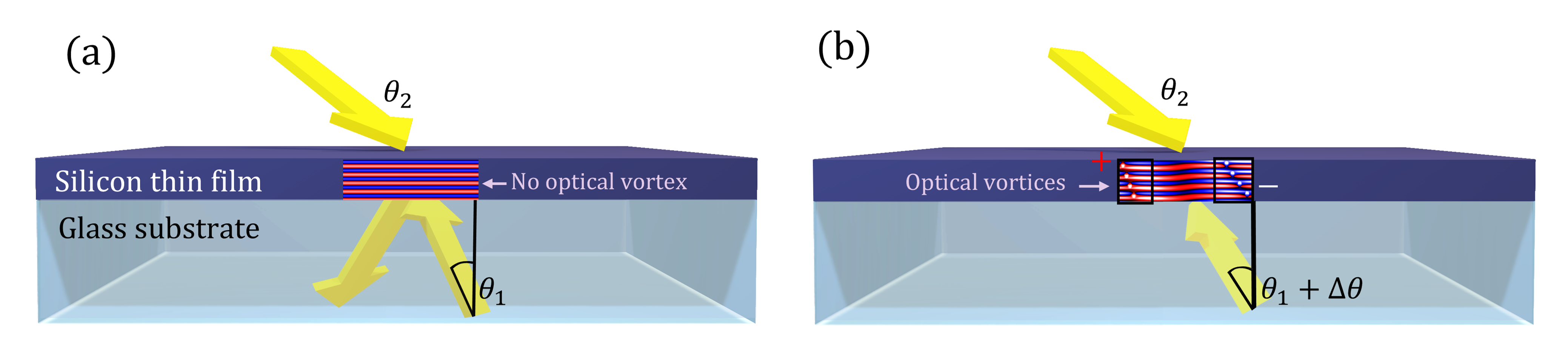}
\caption{Schematic of a planar optical detector geometry under illumination with two plane waves. One plane wave is incidence from the bottom at an angle ($\theta_1$) and propagates from the glass substrate to the thin film, while the second plane wave is incidence from the top at an angle ($\theta_2$) and propagates from the air to the thin film. The OV generation and annihilation in the Si film driven by a small change in the incidence angle of one of the waves is illustrated by the corresponding optical interference patterns in the Si layer. \textbf{(a)} A specific incidence angle ($\theta_1 = \theta_{1,tar}$; with the phase difference of $35.58^\circ$ between $\theta_1$ and $\theta_2$) corresponds to the near-zero absorptance case due to annihilation of the OVs and formation of a regular periodic optical interference pattern. \textbf{(b)} In contrast, a slight change ($\pm 3^\circ$) in the incidence angle dramatically improves absorption via OV generation.}
\label{Fig.1}
\end{figure*}
In this section, we first investigate angle-selective OV generation by interfering plane waves within a simple 1D thin film geometry. We draw inspiration from the pioneering work of Wolter~\cite{wolter2009concerning}, which predicted the generation of OVs close to a material interface upon excitation with two OAM-free plane waves incident at different angles. Several follow-up works demonstrated local near-field OV generation at the material interfaces under the excitation of the OAM-carrying vortex beams~\cite{lembessis2009surface,dennis2012topological}. However, these beams have to be preliminarily created by the light wavefront shaping with spiral phase plates, carefully designed metasurfaces, or phased array antennas. In contrast, to harness the near-field OV phenomena for light absorption or biosensing, it is important to develop a comprehensive approach to engineer nanophotonic structures to generate localized trapped OVs under the excitation by plane waves. To maximize the effect of OV generation and energy re-circulation on the efficiency of light absorption in thin films, it is also paramount to develop a design strategy to generate multiple OVs distributed within the optical absorbers with high spatial density. 

First, we revisit the most straightforward idea of Wolter's work and investigate the possibility of angular light sensing in a simple 1D structure shown in Fig.~\ref{Fig.1}, which can be achieved by illuminating the structure with two plane waves at varying angles. Multiple interacting plane waves generally create constructive or destructive interference~\cite{faist1997controlling,davuluri2015destructive} depending on their angles and phases, which, in turn, can lead to either absorption enhancement or reduction. At least two waves need to interfere to create conditions to observe optical phase singularities and the associated OVs.
Figure~\ref{Fig.1} illustrates the schematic of our proposed structure, consisting of a 750 nm thick Silicon film on a glass substrate ($n=1.52$). We employ a full-wave Maxwell solver, Meep Finite-Difference Time-Domain (FDTD)~\cite{oskooi2010meep, sullivan2013electromagnetic}, to calculate the optical interference field and the absorptance enhancement for the optical angular detection. Two s-polarized plane waves illuminate the planar absorber at different incidence angles ($\theta_1$ and $\theta_2$) and directions (from the substrate side and from the top, respectively). We chose a 775 nm wavelength compatible with a common infrared lasing system. At this wavelength, the silicon thin film absorption is relatively limited~\cite{aspnes1983dielectric}. As shown in Figs.~\ref{Fig.1}(a,b), OV generation in the detector near field is very sensitive to small variations in the incidence angle of one of the plane waves, e.g., switching from a regular Fabry-Perot-type periodic interference pattern shown in Fig.~\ref{Fig.1}(a) to a more exotic interference field featuring multiple OV pairs as shown in Fig.~\ref{Fig.1}(b).

\begin{figure*}[!htbp]
\centering
\includegraphics[width=1.0\linewidth]{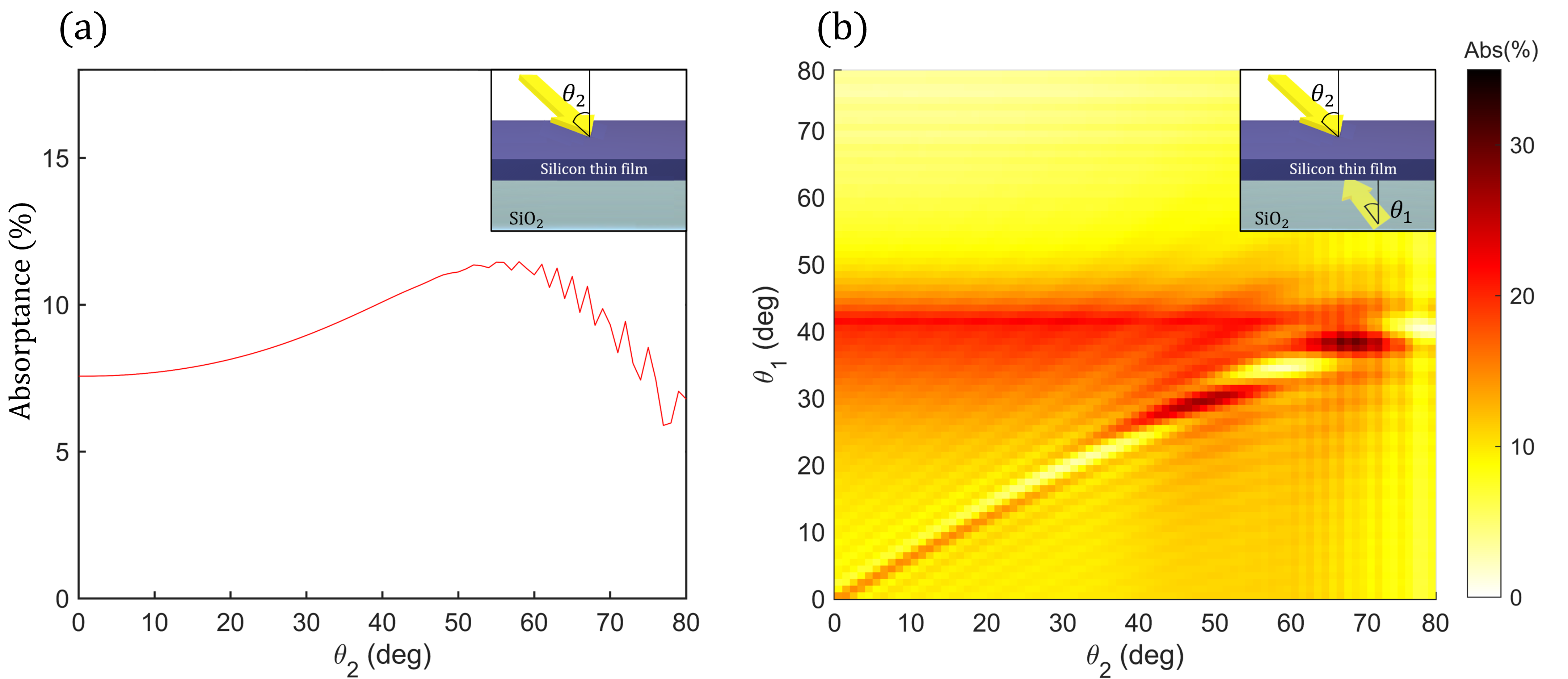}
\caption{The Si layer absorptance as the function of the incidence angle under illumination with either a single or two interfering plane waves with angles of incidence $\theta_1$ and $\theta_2$, each varying from $0^\circ$ to $80^\circ$. \textbf{(a)} The angle-dependent absorptance of a single plane wave propagating from the top side. \textbf{(b)} Two plane waves incident on the absorber from opposite directions yield absorptance characteristics that are highly sensitive to small changes in their incidence angles. This effect can be used to design an optical angular sensor, where we detect the incidence angle of one wave by using the other wave as a probe. The inset figures indicate the incidence angles and the 1D geometry used in the simulations.}
\label{Fig.2}
\end{figure*}
\begin{figure*}[!htbp]
\centering
\includegraphics[width=0.6\linewidth]{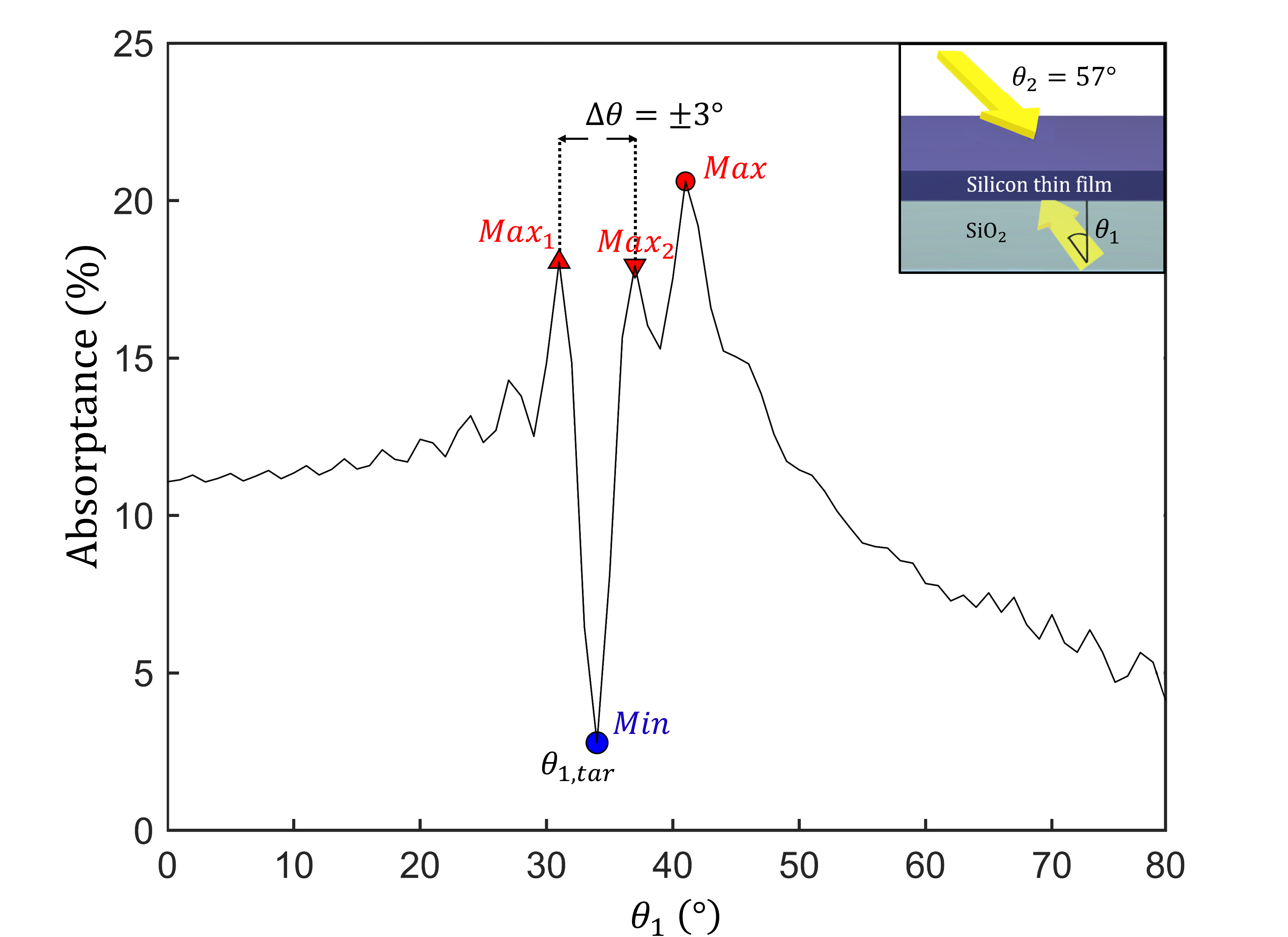}
\caption{Si-film absorptance as a function of the incidence angle of the bottom plane wave ($\theta_1$), while the incidence angle of the plane wave at the top ($\theta_2$) is fixed at $57^\circ$. The inset shows a 1D structure used in the simulations. The minimum absorptance of 3.67\% (indicated as the blue circle) occurs at an incidence angle of $34^\circ$ with the phase difference of $35.58^\circ$ between $\theta_1$ and $\theta_2$. In contrast, a $3.0^\circ$ shift of the incidence angle away from $\theta_{1,tar}$ generates multiple OVs in the 750 nm thick silicon layer, resulting in a 6.05-fold absorptance enhancement.}
\label{Fig.3}
\end{figure*}
\begin{figure*}[!htbp]
\centering
\includegraphics[width=0.85\linewidth]{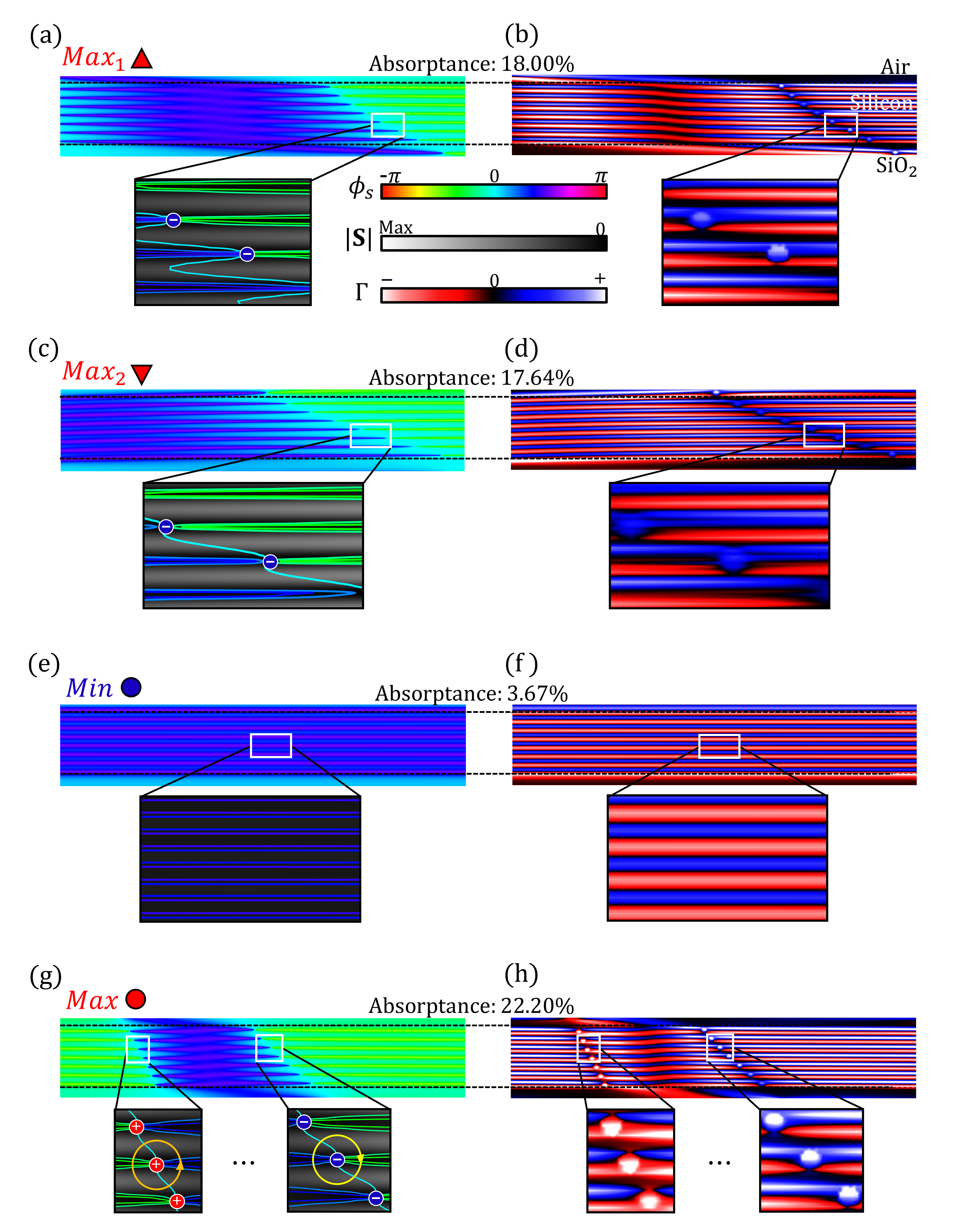}
\caption{The magnitude $\left|\textbf{S}(x,y)\right|$ and the direction $\phi_{s}(x,y)$ of the Poynting vector as well as the power flow circulation parameter $\Gamma(x,y)$ in the silicon thin film under illumination with two plane waves. \textbf{(a), (c), (e), (g)} The optical power flow direction in the silicon thin film at each case corresponding to points ${Max}_1$, ${Max}_2$, ${Min}$, and ${Max}$ in Fig.~\ref{Fig.3}. The inset figures show magnified versions with overlaid equi-phase lines ($\phi_{s}=$const) on top of the heat map of the magnitude of the Poynting vector $\left|\textbf{S}\right|$. \textbf{(b), (d), (f), (h)} The power flow circulation parameter $\Gamma(x,y)$ for each point (${Max}_1$, ${Max}_2$, ${Min}$, and ${Max}$ in Fig.~\ref{Fig.3}). The inset figures show the overlaid geometric equi-phase lines on top of the $\Gamma(x,y)$ heat map. Seven OV pairs with counterclockwise (+) and clockwise (-) phase rotations are observed within a half-wavelength-long (5 \textmu m in length) absorber section in (g,h), while (e,f) panels do not show any OVs. The maximum OV circularity in (g), (h) is 24.02\%.}
\label{Fig.4}
\end{figure*}

Figure~\ref{Fig.2} shows the angle-dependent absorptance of a 750 nm thick silicon thin film. Figure~\ref{Fig.2}(a) shows the angle-dependent film absorptance under illumination with a single plane wave propagating from the top side. The absorptance of the thin film illuminated with a single plane wave varies from 5.89\% to 11.47\% over the $0^\circ$ to $80^\circ$ angular sweep. Figure~\ref{Fig.2}(a) reveals that the planar 1D structure exhibits neither high absorptance nor high angular sensitivity under illumination with a single plane wave. On the other hand, as shown in Fig.~\ref{Fig.2}(b), illumination of the planar absorber with two plane waves launched in opposite directions results in the interference field that exhibits a high sensitivity of absorptance to the angles of incidence of the two waves. Angle-sensitive absorption characteristics provide an opportunity for realizing an efficient angular-selective near-infrared photodetector~\cite{jing2016angular} in an elementary device geometry. Figure~\ref{Fig.3} shows the angular sensitivity of the 1D structure when one of the angles $\theta_2$ is fixed to 57$^\circ$. The absorptance curve shows two local maxima ($\theta_1=31$ and $37^\circ$), one global maximum ($\theta_1=41^\circ$), and one global minimum point ($\theta_1=34^\circ$). For further analysis, we label these four points as ${Max}_1$, ${Max}_2$, ${Max}$, and ${Min}$, respectively. At the ${Min}$ point (which corresponds to the phase difference of $35.58^\circ$ between $\theta_1$ and $\theta_2$), absorptance is only 3.67\%, while the local maxima points (${Max}_1$ and ${Max}_2$) show absorptance of 18.00\% and 17.64\%, respectively, which is more than 4.8 times higher than the absorptance at the ${Min}$ point despite very small angle difference ($\pm 3^\circ$) between these points. Fig.~\ref{Fig.3} demonstrates that our simple planar optical detector design shows a smaller angular half-width-half-maximum (HWHM) value ($1.5^\circ$) than previously reported HWHM values for angle-selective silicon-on-insulator ($1.7^\circ$)~\cite{nagarajan2020angle} and infrared photodetector ($15^\circ$)~\cite{jing2016angular} device designs. These observations suggest that our approach may lead to the realization of the highly sensitive optical angular sensor using a simple 1D structure. 

To reveal the correlation between the enhancement and suppression of the thin-film absorptance observed in Fig.~\ref{Fig.3} and the generation and annihilation of OVs in the absorber near field, we calculate and plot in Fig.~\ref{Fig.4} the spatial distributions of the magnitude and direction of the Poynting vector of the two-wave interference field at the wavelengths corresponding to ${Max}_1$, ${Max}_2$, ${Max}$, and ${Min}$ points. The time-averaged Poynting vector can be calculated with the following equation:
\begin{equation} 
\begin{aligned}
\textbf{S}(x,y)&=\left(S_x(x,y), S_y(x,y)\right)\\
 &=\frac{1}{2}\textrm{Re}\left[\hat{\textbf{E}}(x,y)\times\hat{\textbf{H}}^*(x,y) \right],
\end{aligned}
\label{Poynting vector}
\end{equation}
where $\hat{\textbf{E}}$ and $\hat{\textbf{H}}$ are the complex amplitudes of the electromagnetic field. The direction of the optical energy power flow at each point in space can be characterized by calculating the orientation of the Poynting vector, defined by the angle $\phi_s$ it makes with the x-axis:
\begin{equation} 
\phi_s=\arcsin\left(\frac{S_y(x,y)}{\left|\textbf{S}(x,y)\right|}\right), \phi_s=\arccos\left(\frac{S_x(x,y)}{\left|\textbf{S}(x,y)\right|}\right).
\label{Geometric phase}
\end{equation}
In turn, $\left|\textbf{S}(x,y)\right|$ is the magnitude of the Poynting vector, which defines the strength of the local power flow.
We also introduce a new parameter to characterize and compare the circularity and strength of local optical power flow at each point of the interference field:
\begin{equation} 
\Gamma(x,y)=\int\nolimits_0^{2\pi}\textbf{S}(x',y')\cdot \hat{\textbf{a}}_\phi d\phi.
\label{QOV}
\end{equation}

Here, the value of $\Gamma(x,y)$ at each point in space $x$, $y$ is calculated by integrating a scalar product of the Poynting vector and a unit vector $\hat{\textbf{a}}_\phi$ in a cylindrical coordinate system over an infinitesimal circle $x'=x+R\cos\phi$, $y'=x+R\sin\phi$ surrounding this point. Thus, this parameter is maximized around the points in space where local areas of strong power flow circulation are present, i.e., around OVs. It should be noted that while OVs are local topological features of the optical field characterized by a distinct discrete topological charge, the spatial extent of the circulating power flow region surrounding the phase singularity in the center of OV depends on the presence of other OVs in their vicinity. $\Gamma(x,y)$ parameter allows to quantify this property of the optical power flow, and provides more information than the Poynting vector angular phase $\phi_s$ distribution alone. The strength and circularity of the local power flow are expected to translate into absorption enhancement as it both extends the effective light-matter interaction distance and boosts the local electromagnetic field intensity. Furthermore, a normalized $\Gamma(x,y)$, OV circularity, can be defined by normalizing $\Gamma(x,y)$ by its maximum value, which is reached if the local energy power flow forms a perfect circle. 

Figure~\ref{Fig.4} shows the magnitude and the direction of the Poynting vector as well as the new power flow circulation parameter ($\Gamma(x,y)$) in the near field of the 750 nm thick silicon film for each scenario (${Max}_1$, ${Max}_2$, ${Min}$, and ${Max}$) indicated in Fig.~\ref{Fig.3}. The images in the insets of Fig.~\ref{Fig.4}(a), (c), and (g) illustrate how destructive interference of two plane waves creates multiple OVs within the silicon layer, translating into greater absorptance. In comparison, as shown in Fig.~\ref{Fig.4}(e), constructive interference of two opposite plane waves generates a typical periodic intensity pattern but does not lead to the formation of OVs and, counter-intuitively, leads to low absorptance (3.67\% in the ${Min}$ case). On the other hand, the optimum absorptance (22.20\%), depicted in Fig.~\ref{Fig.4}(g) and (h), is driven by strong local circulatory optical power flow in the form of seven pairs of counterclockwise ($+$) and clockwise ($-$) OVs, which are characterized by high values of $\Gamma(x,y)$. In turn, the optical interference field corresponding to the other local absorption maxima, depicted in Fig.~\ref{Fig.4}(a)-(d), (18.00\%, 17.64\%) also feature multiple OV pairs, which, however, have lower circulation parameter values as compared to the ${Max}$ case. These results confirm that the formation of real-space OVs underlies the absorption spectral features in thin films, and the circulation parameter can serve as a good measure for the resonant absorptance enhancement. The maximum OV circularity in Fig.~\ref{Fig.4}(g) is 24.04\%.
\begin{figure*}[!htbp]
\centering
\includegraphics[width=0.6\linewidth]{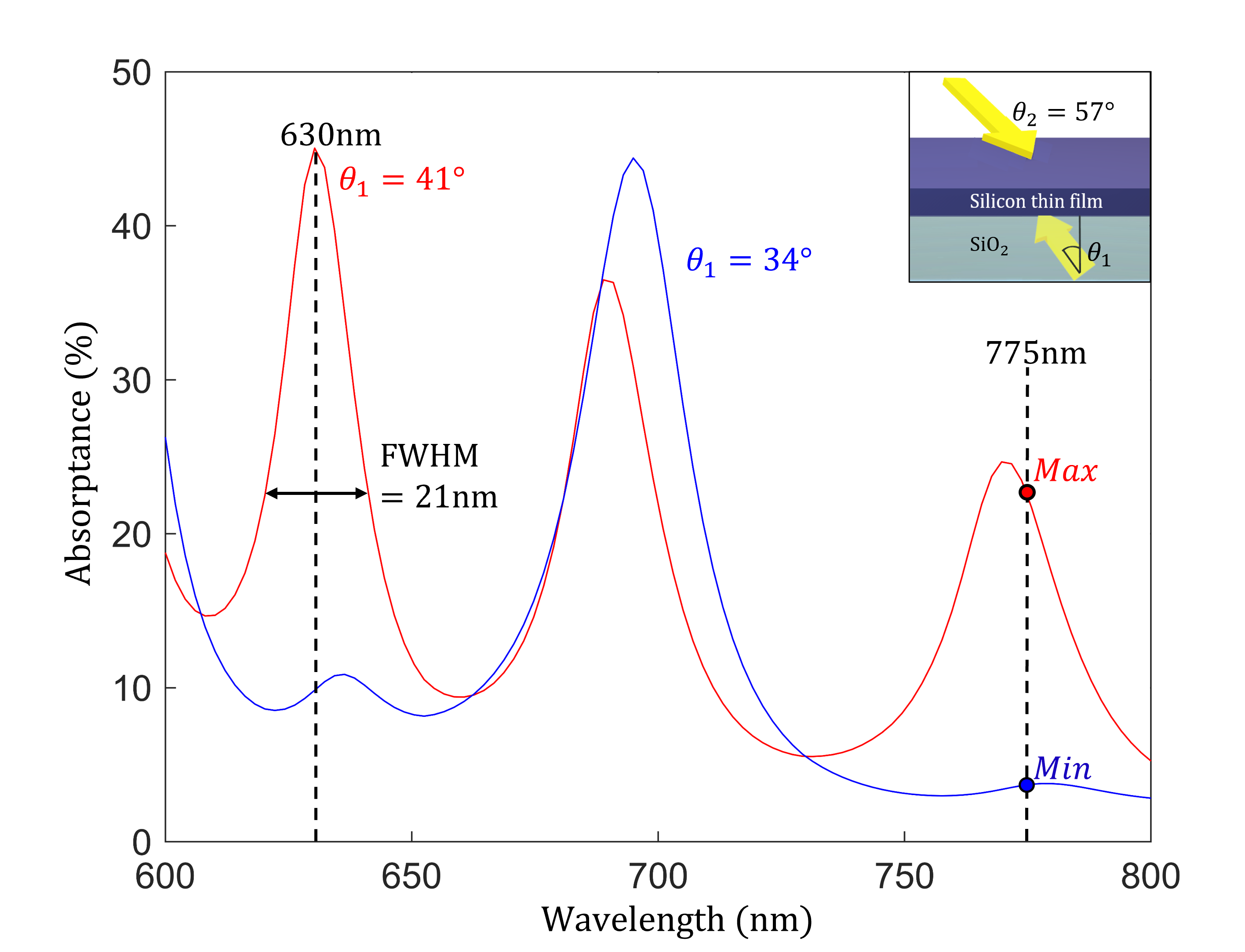}
\caption{Absorptance spectra of the same planar structure as in Figs.~\ref{Fig.3}-\ref{Fig.4} over the wavelength range from 600 to 800 nm for the two angles of incidence of the upward-propagating probe wave corresponding to the ${Min}$ and ${Max}$ cases. The blue curve represents the ${Min}$ ($\theta_1=34^\circ$ ), and the red curve represents the ${Max}$ ($\theta_1=41^\circ$) case, while the downward-propagating plane wave arrives at an angle of $\theta_2=57^\circ$. The absorptance contrast ratio between ${Min}$ and ${Max}$ is maximized at the wavelengths of 775 nm (shown in Figs.~\ref{Fig.3}-\ref{Fig.4}) and 630 nm. The absorptance peak at 630 nm exhibits a relatively narrow FWHM (21 nm), which could be leveraged for the design of a wavelength-selective and angle-selective sensor~\cite{ogawa2012wavelength}.}
\label{Fig.5}
\end{figure*}

Figure~\ref{Fig.5} shows simulated absorptance spectra for two different incidence angles (corresponding to the cases labeled as $\theta_{Min}=34^\circ$ and $\theta_{Max}=41^\circ$ in Fig.~\ref{Fig.3}). At the wavelength of 775 nm, the probe plane wave arriving from the bottom at $41^\circ$ interferes with the downward-propagating plane wave, creating OVs and yielding a total absorptance of 20.61\%. In contrast, the interference of the two waves when the probe wave arrives at the angle of $34^\circ$ yields only 3.67\% absorptance. The contrast absorptance ratio at 775 nm is 6.05, while a similar high-contrast absorptance ratio is observed as 4.57 at the wavelength of 630 nm. When the incidence angle ($\theta_1$) of the probe plane wave is $41^\circ$, the full-width-half-maximum (FWHM) is estimated as 21 nm at the 630 nm wavelength. The high-absorptance-contrast wavelength position and width can further be tuned to overlap with a specific laser emission spectrum by the judicious choice of absorber thickness.
Both angular and wavelength sensitivity of the absorptance to the probe wave parameters, which are underlied by the OV generation in the interference field, can be leveraged to design angle-sensitive detectors~\cite{nagarajan2020angle,li2022dual}, wavelength-sensitive detectors~\cite{cleary2013long,sreekanth2018ge2sb2te5,wu2021bandwidth,li2022dual}, and color filters~\cite{li2015large}.

However, while this approach enables efficient absorption suppression at the ${Min}$ point, the level of absorptance at the maximum absorption points is only around 20\%. Furthermore, aligning the two beams and precisely controlling their phases could still be challenging. Therefore, in the following sections, we consider the case of a single wave incidence, and instead of interfering two waves, we optimize the absorptance in a thin Si film by engineering nano-structured metasurface coatings made from lossless dielectrics~\cite{chung2021inverse} using the inverse design method~\cite{pestourie2018inverse, mansouree2020multifunctional, li2022empowering}. This allows us to demonstrate a dramatic absorption enhancement in a much thinner silicon absorber (100 nm thick). As we show in the following, our design approach allows achieving perfect absorption in subwavelength-thick films by decoupling the photon and electron propagating lengths in composite absorber structures, which should further lead to increased performance of infrared detectors~\cite{atwater2010plasmonics}.

\section{Inverse design of ultrathin perfect absorbers}
The inverse design, also known as adjoint optimization, enables large-scale optimization of photonic structures, such as metalenses~\cite{molesky2018inverse,chung2020high,lim2021high,chung2023inverse}, solar cells~\cite{miller2012photonic, ganapati2013light,molesky2018inverse}, bio-sensors~\cite{chung2021inverse,chung2022inverse}, and OV beam emitters~\cite{doi:10.1021/acsphotonics.2c01007,otte2023tunable}. The inverse design algorithm is based on calculating derivatives of the chosen figure of merit (FoM) induced by the change of the permittivity $\delta\text{FoM}/\delta\epsilon(\textbf{x})$ with only two simulation runs (a direct and an adjoint simulation). Specifically, a slight change of the permittivity in the designable region can be approximated as a superposition of a simulation with no change and a simulation with a dipole excitation at the location of the permittivity change. Then, the Lorentz reciprocity principle is used to replace many simulations of dipoles at different positions within the designable region with a single dipole simulation at the detector location~\cite{miller2012photonic}. This process simplifies calculation of derivatives of FoM with respect to the change of permittivities ($\delta\text{FoM}/\delta\epsilon(\textbf{x})$) with the dot product of electric field ($\textbf{E}_{Dir}$) from the incidence wave and the adjoint field ($\textbf{E}_{Adj}$) from the dipole source. The back-propagating field of the adjoint source is given by $\textbf{J}_{Adj}=-i\omega\textbf{P}=-i\omega(\delta\text{FoM}/\delta\textbf{E})$, where $\textbf{P}$ is a polarization density. This technique enables us to solve complicated photonic problems with many design parameters within feasible computational time. 

Prior studies indicate that a combination of periodicity and randomness in the absorber surface design is vital to achieving greater absorptance in dielectric thin films~\cite{ganapati2013light,chung2014time}. In this work, we employ a metasurface design periodicity of 1.5 \textmu m and define FoM as the maximum of electric field intensity ($|\textbf{E}(x_{0},y_{0})|^2$) at the center of 100 nm thick silicon film at a 775 nm wavelength. We have tested multiple field-maximization points and different FoMs, and our results demonstrate that maximizing the field intensity at a well-chosen periodic distance (1.5 \textmu m) can yield a design exhibiting a perfect absorption in the 100 nm thick silicon film. We employ Meep FDTD~\cite{oskooi2010meep} combined with a pre-qualified in-house inverse design algorithm~\cite{chung2021inverse,chung2023inverse}.
\begin{figure*}[!htbp]
\centering
\includegraphics[width=0.6\linewidth]{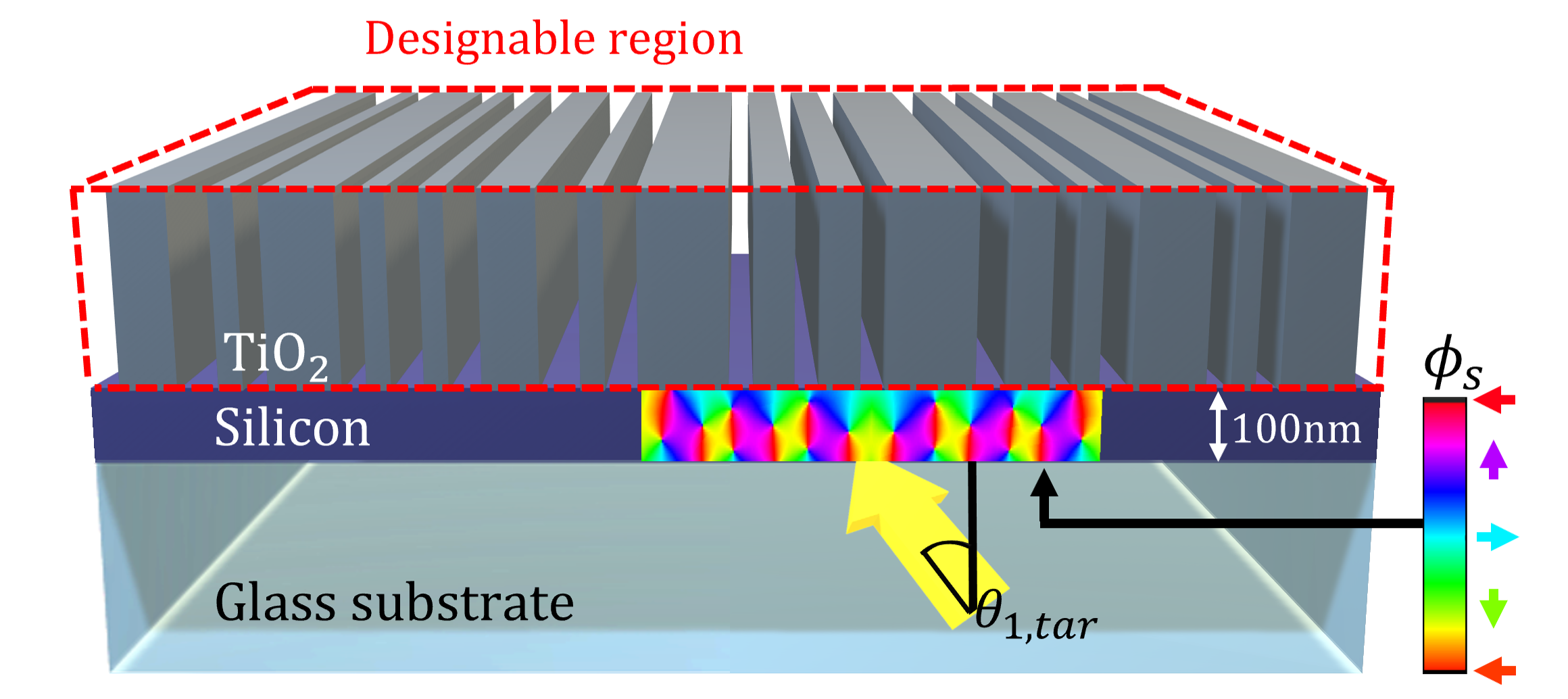}
\caption{Schematic of the inverse design problem. Through inverse design, we optimize the TiO$_2$ layer geometry to maximize the OV-driven absorptance in a 100 nm thick silicon film under the illumination with a plane wave incident at the absorber from the glass layer side at angle $\theta_{1,tar}$. Multiple OV pairs form within the thin film at frequencies corresponding to the perfect absorption condition. The side surface plot (accompanied by the color bar and the arrow on the right-hand side) shows the Poynting vector phase distribution ($\phi_{s}$) within the Si film defined by Eq~\ref{Geometric phase}.}
\label{Fig.6}
\end{figure*}

Figure~\ref{Fig.6} shows a schematic illustrating a design of a structure achieving perfect absorption in a planar Si thin film driven by the OV generation in the interference near-field, which is optimized through the inverse design method. We consider a 100 nm thick silicon layer on a glass substrate illuminated with a 775 nm-wavelength plane wave. At this wavelength, silicon has a low absorption coefficient of 1.10$\times 10^3$ ($\text{cm}^{-1}$)~\cite{green1995optical}, resulting in 3.67\% single-pass absorptance in a planar layer. The designable region consisting of TiO$_2$ layer is placed above the silicon film. We first use the inverse design process to optimize the geometry of the TiO$_2$ layer without any constraints (as a free-form metasurface) and then apply additional minimum feature size constraints to obtain a lithography-compatible grating-type structure. A single s-polarized plane wave illuminates the structure from the glass side, and we show in the following that the interference field generated by its scattering on the optimum inverse-designed TiO$_2$ metasurface features multiple coupled OV pairs, which drive light recycling and absorptance in the underlying planar Si thin film.

First, we optimize the designable region with a free-form TiO$_2$ metasurface geometry. In this simulation, the density of TiO$_2$ is allowed to vary between 0 and 1 at every location in the designable region, known as the topology optimization approach. Then, penalty functions~\cite{neves2002topology} are added to the FoM function to impose a binary-material constraint (ultimately, allowing only 0 and 1 density values of material at each location). 

Later in this section, we also impose the restriction of constant material density in the z-direction, which ensures the structure manufacturability via traditional lithography methods~\cite{chen2015nanofabrication}. The initial density parameters are selected randomly within a range of 0.45 to 0.55.
The free-form-based optimization can provide insight regarding a computational upper limit of the FoM for the given optimization problem~\cite{shim2020maximal, chung2022computational}. A free-form design can have numerous degrees of freedom depending on the grid spacing chosen for a simulation. Therefore, an optimized free-form structure can potentially yield a design with the FoM reaching a physical bound of a given structure~\cite{shim2020maximal}.
\begin{figure*}[!htbp]
\centering
\includegraphics[width=1.0\linewidth]{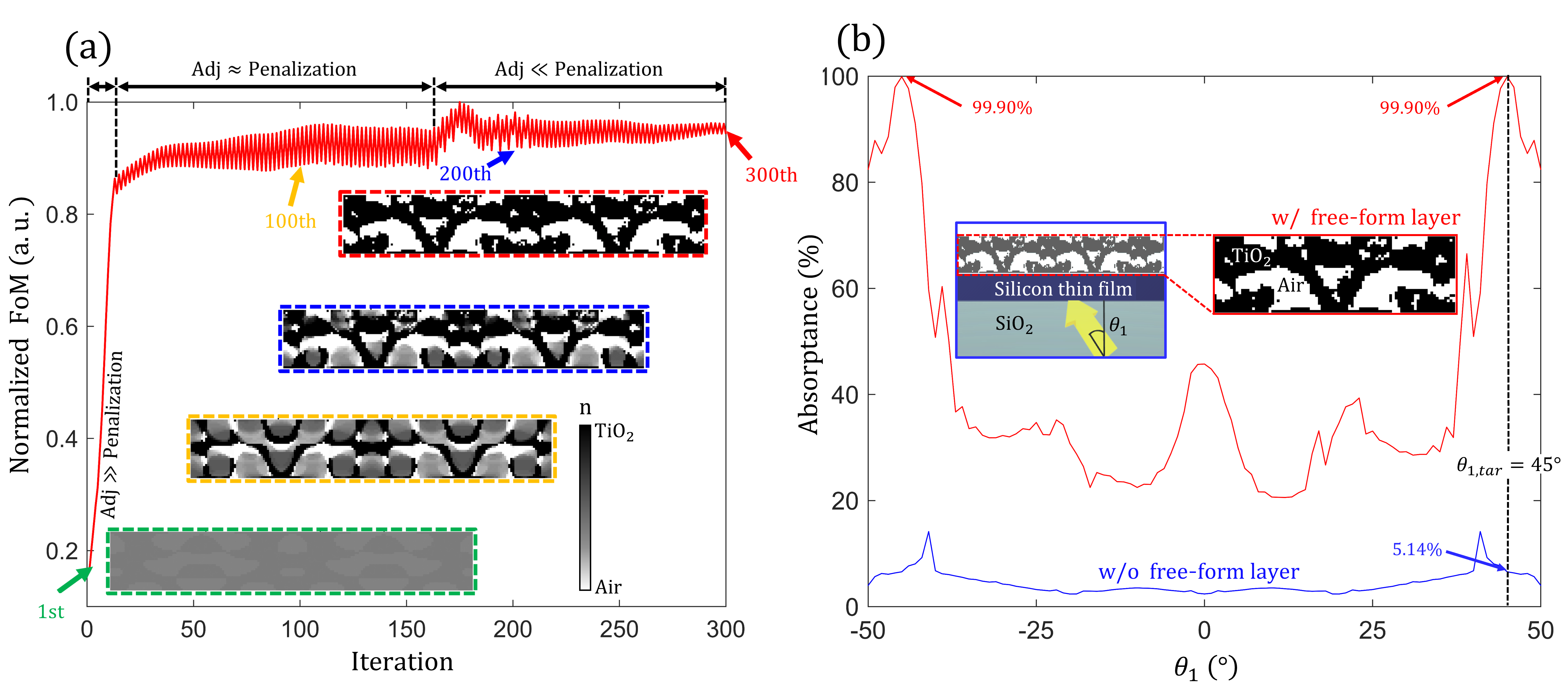}
\caption{The inverse design procedure for optimizing a TiO$_2$ free-form structure. \textbf{(a)} The FoM is defined as the maximum of the electric field intensity at the center of the thin film when a 775 nm-wavelength s-polarized plane wave is incident at the target angle ($\theta_{1,tar}=45^\circ $). The red curve indicates the evolution of the FoM, while the four insets show the changes in the design parameters. The penalization factor gradually increases with the iteration number. Then it starts to exceed the gradients of FoM after the $144_{\text{th}}$ iteration, which enforces grayscale design parameters to be binarized to TiO$_2$ and air. \textbf{(b)} Absorptance of the inverse-designed structure as a function of the incidence angle of the plane wave ($\theta_1$). The blue curve represents the single-pass absorptance, and the red curve corresponds to the absorptance of the inverse-designed structure, which shows perfect absorption (99.90\%) at $\theta_1=45^\circ$. This constitutes a 19.44-fold absorptance improvement compared to the single-pass absorptance at the same angle of incidence. The inset figure shows an optimized structure and its surrounding materials.}
\label{Fig.7}
\end{figure*}

Figure~\ref{Fig.7}(a) shows several inverse design iterations exhibiting progressively increasing values of the normalized FoM. At the initial iterations, adjoint derivatives with respect to the change of permittivity ($\delta\text{FoM}/\delta\epsilon(\textbf{x})$) dominate the optimization process. Then, the penalization factor converts grayscale permittivity values to binarized permittivity values (corresponding to those of TiO$_2$ and air). The penalization factor gradually increases over the inverse design iteration until the design parameters are fully binarized, as shown in the red inset figure in Fig.~\ref{Fig.7}(a). Figure~\ref{Fig.7}(b) shows the absorptance of the structure as a function of the plane wave incidence angle. The inverse-designed structure exhibits perfect absorption (99.90\%) at the target incidence angle ($-45^\circ$, $45^\circ$). It is a 19.44-fold enhancement compared to the single-pass absorptance at the target angle, as shown with the blue curve in Fig.~\ref{Fig.7}(b).
\begin{figure*}[!htbp]
\centering
\includegraphics[width=1.0\linewidth]{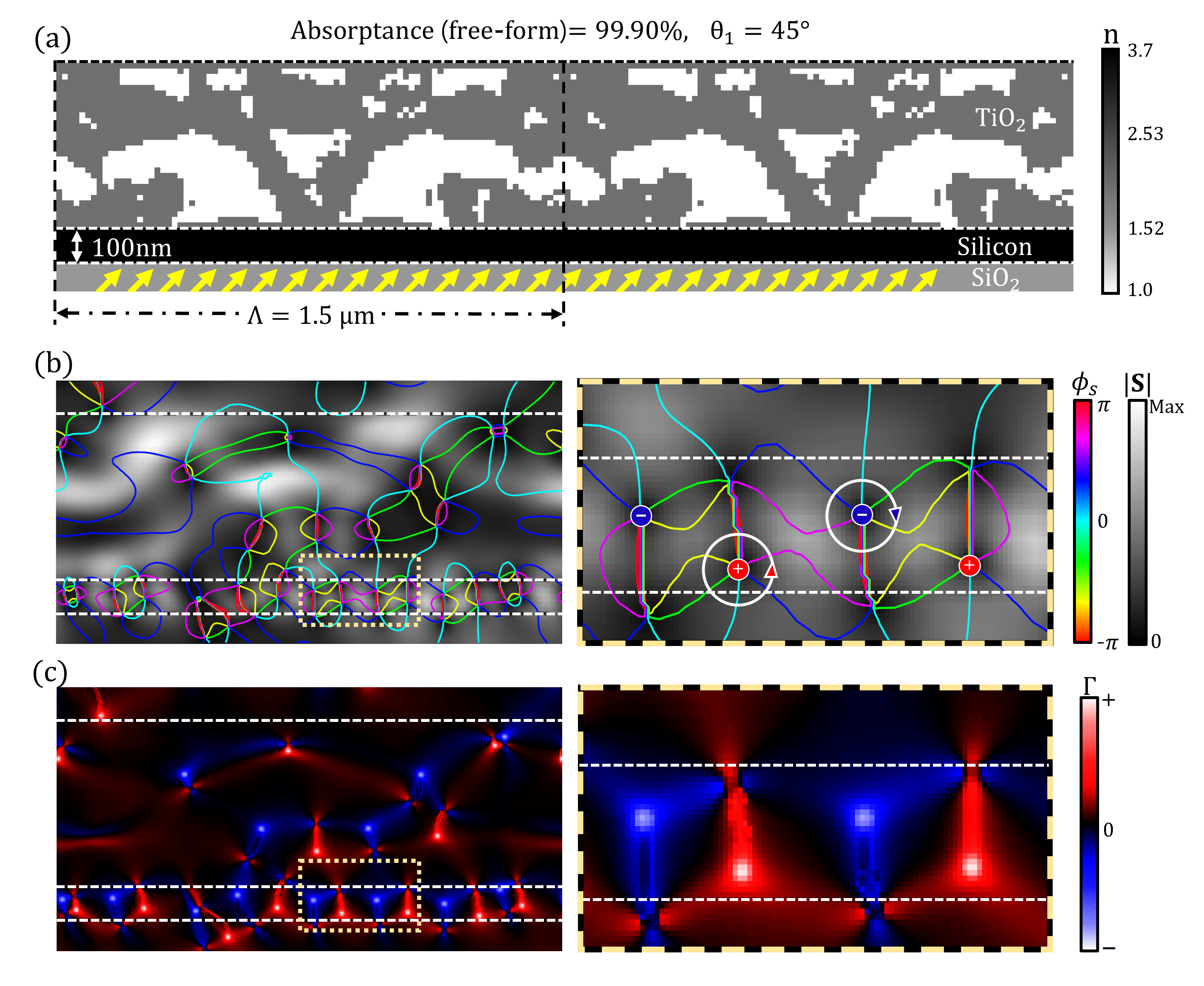}
\caption{The OV generation inside the inverse-designed TiO$_2$ free-form structure at the perfect absorption. \textbf{(a)} The geometry of the optimized free-form TiO$_2$ structure with a 100 nm thick silicon absorber layer and a SiO$_2$ substrate. The design periodicity is chosen as 1.5 \textmu m, and the plane wave is incident a $45^\circ$ angle. The optimized design shows 99.90\% absorptance, which is a 19.44-fold improvement compared to the absorptance of a planar 100 nm thick silicon film on SiO$_2$ substrate illuminated with a plane wave at the same incidence angle and frequency. \textbf{(b)} The phase $\phi_{s}(x,y)$ and the magnitude $\left|\textbf{S}(x,y)\right|$ of the Poynting vector. The colored contour curves are the equi-phase lines ($\phi_{s}=$const) overlaid above the heat map showing the magnitude of the Poynting vector. The zoomed-in version of the plot in the area inside the yellow dashed box is shown on the right-hand side and illustrates clockwise and counter-clockwise OV pairs formed in the absorber layer. \textbf{(c)} The OV circulation parameter $\Gamma(x,y)$ distribution in the inverse-designed free-form geometry. The area inside the yellow dashed box is shown with higher resolution on the right-hand side to highlight the OV circulation character. Seven pairs of OVs are observed in the 100 nm thick silicon layer, with an average circularity of 88.18\% and a maximum of 95.37\%.}
\label{Fig.8}
\end{figure*}

Figure~\ref{Fig.8}(a) illustrates the optical power flow patterns inside the optimized inverse-designed free-form TiO$_2$ geometry under the illumination with a plane wave at a $45^\circ$ incidence angle. The OV pairs formed near the thin-film Si absorber are revealed in Fig.~\ref{Fig.8}(b) by plotting the equi-phase lines ($\phi_{s}=$const) overlaid on top of the heat map showing the magnitude of the Poynting vector $\left|\textbf{S}(x,y)\right|$. Finally, Fig.~\ref{Fig.8}(c) shows the spatial distribution of the OV circulation parameter ($\Gamma(x,y)$). Both the Poynting vector and the circulation parameter have been plotted over a single periodic unit cell (of 1.5 \textmu m in width) of the optimized structure. The Poynting vector distribution in Fig.~\ref{Fig.8}(b) shows highly localized optical energy flow and circulation near the 100 nm thick silicon layer, which leads to its perfect absorption (99.90\%). The equi-phase lines of the Poynting vector reveal local areas of clockwise or counter-clockwise circulating optical power flow corresponding to the OV formation. The inset of Fig.~\ref{Fig.8}(b) (shown on the right) offers a close-up view of the OV pairs enclosed in the dashed box area marked in Fig.~\ref{Fig.8}(b) (left panel). The OV circulation parameter $\Gamma(x,y)$ is plotted in Fig.~\ref{Fig.8}(c) over the same area as in Fig.~\ref{Fig.8}(b). The inset of Fig.~\ref{Fig.8}(c) (shown on the right) zooms into the OVs with high circulation parameter values, which are enclosed in the dashed box area in Fig.~\ref{Fig.8} (c) (left panel). The average circularity in this spatial area is calculated to be 88.18\%, and the maximum value of circularity is 95.37\%. The images in Fig.~\ref{Fig.8}, as well as the perfect absorptance value, confirm that the incident plane wave is fully converted to circulating optical flow within the thin film as a result of the near-field optical interference in the optimized inverse-designed TiO$_2$ structure. 

While the free-form-based optimization converged to a structure with the optimum FoM, experimental realization of the 2D free-form structure shown in Fig.~\ref{Fig.8}(a) is hardly feasible. Thus, to design a structure that can be manufactured by conventional lithographic techniques, we constrain our designable region to a 2D grating structure featuring a translational symmetry in the direction perpendicular to its cross-sectional view. 
\begin{figure*}[!htbp]
\centering
\includegraphics[width=1.0\linewidth]{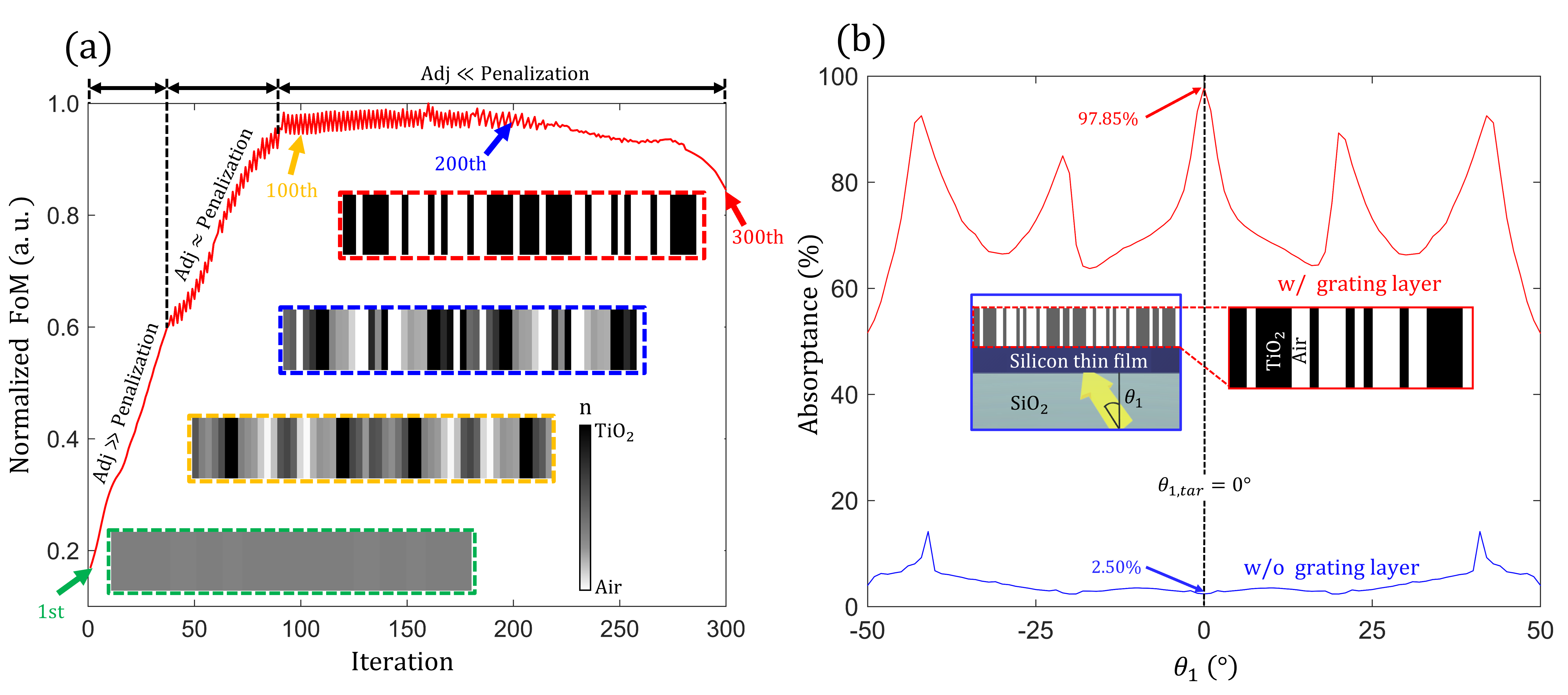}
\caption{The inverse design procedure for optimizing a TiO$_2$ grating. \textbf{(a)} Normalized FoM values versus the inverse design iteration number. The red curve illustrates the evolution of the FoM, while the four subplots show the corresponding changes in the design parameters. Initially, the penalization factor gradually increases with the increased number of iterations. Eventually, it exceeds the gradients of FoM after the $90_{\text{th}}$ iteration, which enforces grayscale design parameters to be binarized to TiO$_2$ and air. The four subplots show the changes in the design parameters, where the aspect ratio of the design parameter is 10. \textbf{(b)} The absorptance as a function of the incidence angle of the plane wave ($\theta_1$). The blue curve represents the single-pass absorptance (without grating). The red curve indicates the absorptance of the inverse-designed grating structure, which shows perfect absorption (97.85\%) at $\theta_{1,tar}=0^\circ $. This constitutes an 18-fold absorptance improvement compared to the single-pass absorptance level. The inset figure shows the optimized grating metasurface structure on top of the planar Si absorber and a glass substrate}
\label{Fig.9}
\end{figure*}
\begin{figure*}[!htbp]
\centering
\includegraphics[width=1.0\linewidth]{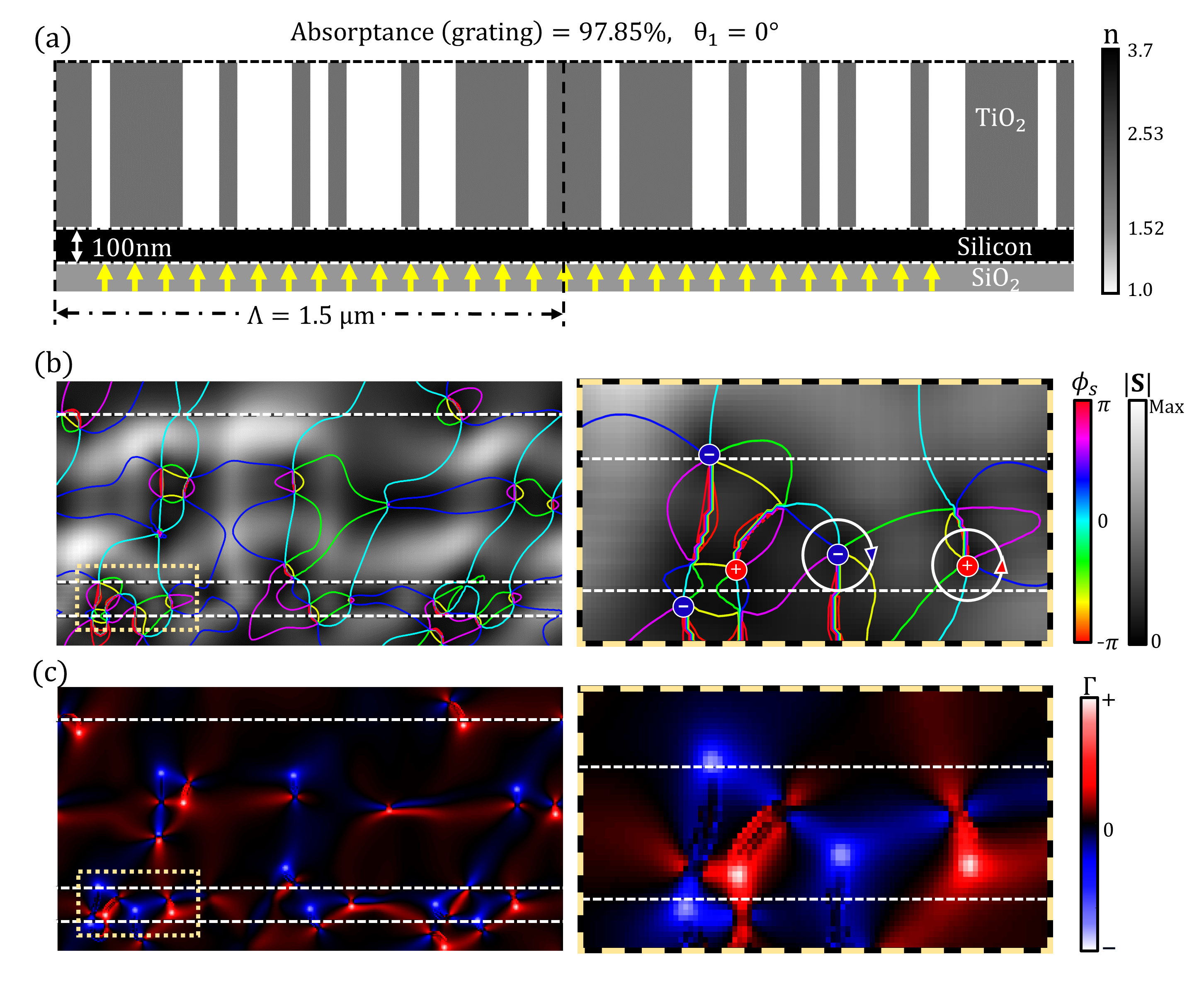}
\caption{OV generation in the optimum inverse-designed TiO$_2$ grating structure. \textbf{(a)} The optimized grating TiO$_2$ structure with a 100 nm thick silicon absorber layer and a SiO$_2$ substrate. The design periodicity is chosen as 1.5 \textmu m, and the incident plane wave arrives normal to the interface (i.e., at a $0^\circ$ angle). The optimized design exhibits 97.85\% absorptance, which is an 18-fold improvement over that of a bare planar 100 nm thick silicon layer on a SiO$_2$ substrate. \textbf{(b)} The phase $\phi_{s}(x,y)$ and the magnitude $\left|\textbf{S}(x,y)\right|$ of the Poynting vector. The colored contour curves are the equi-phase lines ($\phi_{s}=$const) overlaid over the heat map showing the magnitude of the Poynting vector. The area inside the yellow dashed box is zoomed into where OVs are the most densely packed to show clockwise and counter-clockwise OV pairs (right panel). \textbf{(c)} The OV circulation parameter $\Gamma(x,y)$ distribution in the inverse-designed grating geometry. The area inside the yellow dashed box is zoomed into the same area to highlight high-circulation OVs (right panel). Four pairs of high-circulation OVs are observed in a 100 nm thick silicon layer, where their average circularity is 89.51\%, and the maximum circularity is 96.14\%.}
\label{Fig.10}
\end{figure*}

In this work, we cap the maximum allowed grating height at 500 nm, while the minimum width of the smallest grating feature is capped at 50 nm, which together leads to an aspect ratio of 10. Such 2D TiO$_2$ gratings on SiO$_2$ substrates can be fabricated by electron beam lithography~\cite{zeitner2012high, chen2015nanofabrication, li2016high, chung2023inverse} and nanoimprint lithography~\cite{guo2007nanoimprint, chen2015applications, gu2022fabrication}. The inverse design of the grating structure starts with defining grayscale permittivity values in the designable region, as shown in the inset enclosed by the green dashed line in Fig.~\ref{Fig.9}(a). The permittivities of the designable region gradually evolve to achieve greater FoM. Once the FoM saturates, grayscale permittivities are reduced to the binary values, corresponding to either TiO$_2$ or air. As discussed previously, the penalization factor gradually increases over the inverse design iteration until the design parameters are fully binarized, as shown in a red inset in Fig.~\ref{Fig.9}(a). Figure~\ref{Fig.9}(b) shows the absorptance as a function of the incidence angle. The inverse-designed TiO$_2$ grating enables perfect absorption (97.85\%) in the planar Si film at the target incidence angle ($0^\circ$). 97.85\% absorptance constitutes an 18-fold enhancement as compared to the single-pass absorptance at the target angle (see blue curve in Fig.~\ref{Fig.9}(b)).

Figure~\ref{Fig.10}(a) shows the optimum inverse-designed TiO$_2$ grating geometry that achieves perfect absorption under normal incidence. The OV profiles near the thin-film absorber layer are shown in Fig.~\ref{Fig.10}(b), where the equi-phase lines ($\phi_{s}=const$) are plotted on top of the heat map of the magnitude of the Poynting vector $\left|\textbf{S}(x,y)\right|$. Fig.~\ref{Fig.10}(c) illustrates the circulation parameter $\Gamma(x,y)$ distribution over the width (1.5 \textmu m) of one periodic unit cell of the optimized structure. The Poynting vector distribution in Fig.~\ref{Fig.10}(b) shows highly localized optical energy circulation near the 100 nm thick silicon absorber layer, which leads to 97.85\% absorptance. The equi-phase lines of the Poynting vector reveal areas of clockwise or counter-clockwise circulating optical flow around the phase singularities. The inset of Fig.~\ref{Fig.10}(b) (shown on the right) visualizes OV pairs confined in the dashed box area in Fig.~\ref{Fig.10}(b) (left panel). The OV circulation parameter $\Gamma(x,y)$ plotted over the same area as in Fig.~\ref{Fig.10}(b) is shown in Fig.~\ref{Fig.10}(c), and helps to visualize a series of high-circulation OVs inside a 100 nm thick planar silicon absorber. The inset of Fig.~\ref{Fig.10}(c) (shown on the right) zooms into the OVs confined in the dashed box area in Fig.~\ref{Fig.10} (c) (left panel); their average circularity is 89.51\%, and the maximum circularity is 96.14\%. Similar to the free-form metasurface, this proves that the inverse-designed grating structure creates the optical interference field, which enables the coupling of the incident plane wave into circulating optical flow within the thin film. 
In the studies of OV beams, the mode purity is widely used for quantifying OAM beams. Similarly to our OV circularity metric, OV beam mode purity quantifies the degree of the OV power flow deviation from a perfect circle. We find that the OV circulation values in the two optimized structures discussed above are comparable to the values of the OV beam mode purity reported in recent works~\cite{wu2023efficient, liu2023tri, kim2022spontaneous, sroor2020high}.

Both 2D free-form and 2D grating structures exhibit more than 3.53-fold improvement in the OV circularity values compared to the simple 1D structure shown in Fig.~\ref{Fig.4}. This demonstrates the correlation between absorption enhancement and high-circularity localized OV fields and also implies that the inverse design can provide a useful numerical tool for generating OV-rich circulating optical fields for photonic applications.

\begin{figure*}[!htbp]
\centering
\includegraphics[width=0.8\linewidth]{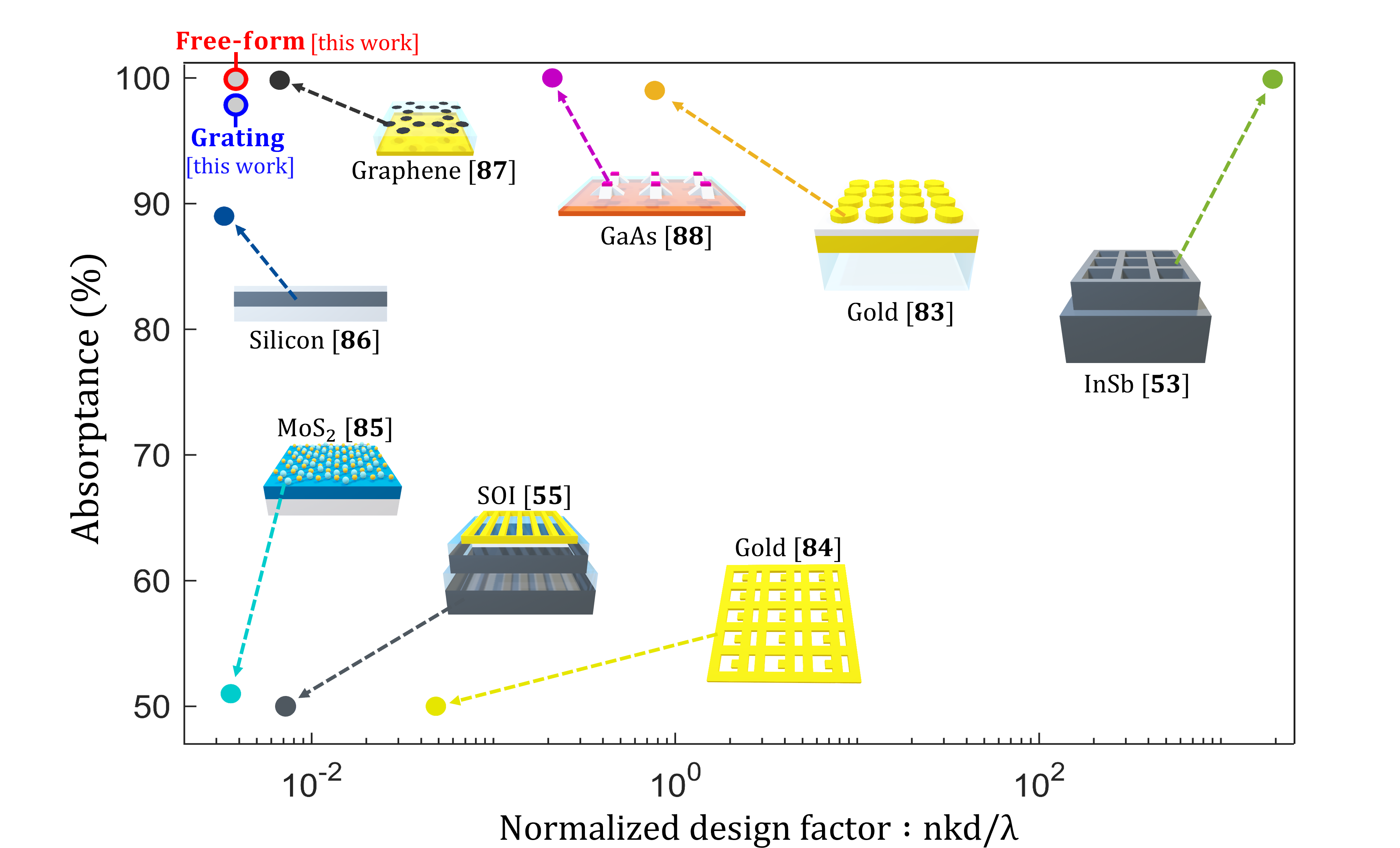}
\caption{Comparison between this work (free-form and grating) and existing works. The normalized design factor is defined as the multiplication of the material thickness `d' and real `n' and imaginary `k' parts of the refractive index divided by the design wavelength `$\lambda$'. The `d/$\lambda$' factor indicates the effective thickness and the imaginary part of the refractive index is highly effective in material absorption. The real part of the refractive index affects the upper limit of the absorption enhancement based on ray-optics limit~\cite{yablonovitch1982statistical, ganapati2013light}. The existing studies regarding thin film absorption include gold~\cite{liu2010infrared, roger2015coherent}, MoS$_2$~\cite{piper2016broadband}, silicon~\cite{yang2020planar, nagarajan2020angle}, graphene~\cite{li2021multi}, InSb~\cite{aspnes1983dielectric} and GaAs~\cite{wang2022plasmonic}. Our designs (free-form, grating) show perfect absorption despite the challenging normalized design factor.}
\label{Fig.11}
\end{figure*}
We compare the performance of the inverse-designed OV-based perfect absorber structures with those from prior works, as shown in Fig.~\ref{Fig.11}. Absorption enhancement depends on (1) the absorber's refractive index (with the maximum enhancement factor of 4n$^2$ known as the Yablonovitch's limit)~\cite{yablonovitch1982statistical}, (2) absorptivity (i.e., the imaginary part of dielectric permittivity) of the material, and (3) the wavelength-normalized absorber thickness. Therefore, we define normalized design factor ($\text{F}$) as $\text{F}=\text{d}nk/\lambda$, where $\text{d}$ is the thickness of the absorber, $\lambda$ is the wavelength of the incident plane wave, $n$ and $k$ are the real and imaginary parts of the absorber refractive index, respectively~\cite{johnson1972optical, beal1979kramers, tikuivsis2023dielectric, aspnes1983dielectric}. 

In Fig.~\ref{Fig.11}, previously proposed ultrathin absorbers are ranked, including structures with gold~\cite{liu2010infrared, roger2015coherent}, MoS$_2$~\cite{piper2016broadband}, silicon~\cite{yang2020planar, nagarajan2020angle}, graphene~\cite{li2021multi}, InSb~\cite{aspnes1983dielectric} and GaAs~\cite{wang2022plasmonic} active layers. Our absorber designs (both the free-form and the grating structures) exhibit perfect absorption despite the challenging normalized design factor. In our designs, a 100 nm thick planar silicon absorber layer is characterized by the refractive index and absorptivity values n= 3.714, k= 0.008 at 775 nm wavelength~\cite{aspnes1983dielectric}, and the wavelength-normalized thickness of 0.129. The inverse-designed free-form structure exhibits 99.90\% absorptance (red dot in Fig.~\ref{Fig.11}), and the inverse-designed grating structure exhibits 97.85\% absorptance (blue dot in Fig.~\ref{Fig.11}). Both designs have a challenging normalized design factor but achieve perfect absorption through near-field OV generation.

\section{Conclusion}
We have demonstrated the correlation between optical vortices (OVs) and absorption enhancement within planar silicon films. Two types of photonic structures have been introduced: a one-dimensional (1D) planar absorber simultaneously illuminated with two incident plane waves at different angles and an inverse-designed two-dimensional (2D) structure illuminated with a single plane wave. 

The former elicits a nearly 4.8-fold absorption enhancement via generation/annihilation of OVs, even with a slight variation ($\pm 3^\circ$ with reference to $\theta_1=34^\circ$) of the incidence angle. This finding shows that absorption enhancement can be achieved by the manipulation of OV and can open new opportunities for engineering angle-sensitive or wavelength-sensitive photodetectors. 

For the latter, we have achieved perfect absorption in the ultrathin film by maximizing the electric field intensity through inverse design. OVs and the resulting light re-circulations are generated from light interference on the optimized dielectric metasurface placed on top of the planar absorber. The inverse-designed free-form TiO$_2$ structure enables 99.90\% absorptance in a 100 nm thick silicon layer at 775 nm wavelength. This is a 19.44-fold absorptance enhancement compared to a single-pass absorption case. We have further simplified the dielectric metasurface geometry to a grating structure that can be fabricated by standard lithographic or nanoimprint techniques. The inverse-designed TiO$_2$ grating structure enables 97.85\% absorptance in the underlying planar Si film, which is an 18-fold improvement over the single-pass case. The OV circularities of optimized structures are a maximum of 95.37\% (free-form) and 96.14\% (grating), and their averages are 88.18\% (free-form) and 89.51\% (grating). Our high OV circularity leads to solid power flow circulation within the absorber, which in turn leads to dramatic absorption enhancement, and these values can be linked to the high OV beam mode purity.

Our work confirms the recent finding that perfect absorption is topological in nature~\cite{tsurimaki2018topological} and reveals that it is accompanied by the generation/annihilation of real-space OVs as well as OVs in the reciprocal (or generalized parameter) space as previously predicted~\cite{liu2023spectral,sakotic2021topological,guo2017topologically, berkhout2019perfect}.


\nocite{*}

\bibliography{apssamp}

\end{document}